\documentclass[aps,prl,twocolumn,showpacs,preprintnumbers,floatfix]{revtex4}
\usepackage{eurosym}
\usepackage{amsmath}
\usepackage{amssymb}
\usepackage{epsfig}

\begin{document}

\title{Extreme driven ion acoustic waves}
\author{L. Friedland}
\email{lazar@mail.huji.ac.il}
\affiliation{Racah Institute of Physics, Hebrew University of Jerusalem, Jerusalem 91904,
Israel}
\author{A. G. Shagalov}
\email{shagalov@imp.uran.ru}
\affiliation{Institute of Metal Physics, Ekaterinburg 620990, Russian Federation}
\affiliation{Ural Federal University, Mira 19, Ekaterinburg 620002, Russian Federation}

\begin{abstract}
Excitation of large amplitude strongly nonlinear ion acoustic waves from a
trivial equilibrium by a chirped frequency drive is discussed. Under certain
conditions, after passage through the linear resonance in this system, the
nonlinearity and the variation of parameters work in tandem to preserve the
phase-locking with the driving wave via excursion of the excited ion
acoustic wave in its parameter space, yielding controlled growth of the wave
amplitude. We study these autoresonant waves via a fully nonlinear warm
fluid model and predict formation of sharply peaked (extreme) ion acoustic
excitations with local ion density significantly exceeding the unperturbed
plasma density. The driven wave amplitude is bounded by the kinetic
wave-breaking, as the local maximum fluid velocity of the wave approaches
the phase velocity of the drive. The Vlasov-Poisson simulations are used to
confirm the results of the fluid model and the Whitham's averaged
variational principle is applied in analyzing evolution of the autoresonant
ion acoustic waves.
\end{abstract}

\pacs{05.45.Yv, 42.65.Tg, 52.35.Sb, 89.75.Kd}
\maketitle

\section{Introduction}

Resonant wave interactions play an important role in plasma applications,
examples being plasma based accelerators \cite{Tajima}, stimulated Raman and
Brillouin scattering in laser driven plasmas \cite{Kruer}, plasma turbulence
\cite{Sagdeev}, etc. These phenomena require phase matching between the
interacting waves, but, frequently, the nonlinear frequency shifts of the
interacting waves destroy the phase matching, limiting the amplitude of the
excitations. Nevertheless, if the parameters of the plasma or of the driving
wave vary slowly in time and/or space, under certain conditions, the
resonant wave interaction may continue despite the nonlinearity due to the
autoresonance effect \cite{Lazar71}, as the interacting waves self-adjust
their amplitudes to stay in a persistent nonlinear resonance. In plasmas,
supporting a variety of nonlinear waves, this phenomenon was studied in the
problem of generation of plasma waves in beat-wave accelerators \cite%
{Lindberg}, in excitation of the diocotron and Bernstein-Green-Kruskal (BGK) modes
\cite{Fajans1, Fajans2, Khain}, and stimulated Raman and Brillouin
scattering \cite{Yaakobi,Williams}. In this paper we exploit autoresonant
wave interactions in the problem of generation of extreme (limited by
kinetic wave breaking) ion-acoustic waves.

The ion-acoustic waves (IAWs) in plasmas were predicted by Tonks and
Langmuir \cite{Tonks} and observed in experiments by \cite{Revans}. Since
these pioneering works, IAWs were studied in many contexts, such as
laser-plasma interactions \cite{Kruer}, ion-acoustic turbulence \cite%
{Bychenkov}, and in dusty \cite{Shukla}, ionospheric \cite{Pavan},
ultra-cold \cite{Castro}, and quantum \cite{Haas} plasmas. Despite of the
importance of the IAWs, their theoretical understanding is still incomplete
in problems involving a combination of nonlinearity, inhomogeneity or time
dependence of the plasma, and kinetic effects. Here, we ask the question of
wether one can resonantly excite and control very large amplitude IAWs via
the autoresonance effect, i.e. by preserving the phase locking between the
driven and driving waves despite the nonlinearity and variation of
parameters. Recently, we have addressed the problem of initiation of
autoresonant excitation of IAWs within a weakly nonlinear fluid model \cite%
{Lazar142}. We have focussed on the case of small ion to electron
temperatures ratio $\sigma ^{2}=T_{i}/T_{e}\ll 1$ to show that by driving
the plasma by a chirped frequency ponderomotive wave passing through the
linear\ ion acoustic resonance, one observes autoresonance in the system if
the driving amplitude $\varepsilon $ exceeds a sharp threshold $\varepsilon
>\varepsilon _{th\text{ }}$. The threshold has the usual autoresonance
scaling $\varepsilon _{th\text{ }}\sim \alpha ^{3/4}$ with the driving
frequency chirp rate $\alpha $ \cite{Fajans1}. We have also seen numerically
\cite{Lazar142} that the amplitude of the autoresonant wave in this setting
can grow well beyond the weakly nonlinear limit. The present work comprises
a fully nonlinear\textit{\ }generalization of the theory going beyond the
usual Korteweg-de-Vries assumption of $k\lambda _{D}\ll 1$ and small
amplitudes \cite{Nicholson}. We will describe the autoresonant growth of the
wave amplitude via chirping the driving frequency, allowing a controllable
approach to the kinetic wave breaking limit.

The scope of the paper will be as follows. In Sec. II we will switch to the
alternative water bag model of IAWs and illustrate excitation of extreme
(maximal amplitude) waves in simulations. In the same section, we will
compare the numerical results of the aforementioned model with the
predictions of the associated kinetic Vlasov-Poisson simulations. In Sec.
III we will describe our fully nonlinear theory of driven autoresonant IAWs.
The approach will be based on the Whitham's average variational principle
\cite{Whitham} using the ideas developed in studying autoresonant excitation
and control of other nonlinear waves \cite{Lazar104,Lazar82}. Finally, Sec.
IV will present our conclusions.

\section{The model and numerical simulations}

\begin{figure}[tp]
\centering \includegraphics[width=8.8cm]{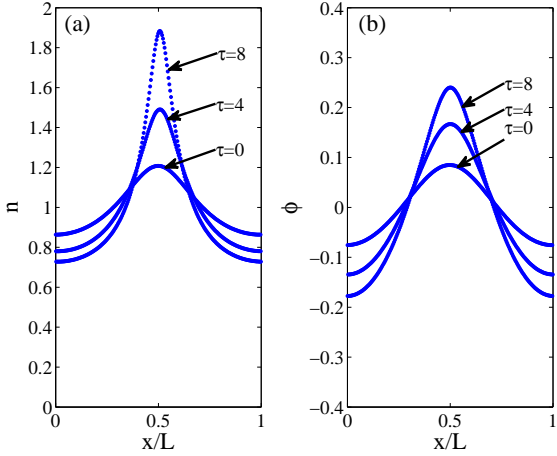}
\caption{(Color online) The spatial profile of the autoresonant ion acoustic
wave (in wave frame) during the excitation process at three different values
of slow time $\protect\tau =0,4,8$. (a) The ion density $n$ and (b) the wave
potential $\protect\phi $.}
\label{fig1}
\end{figure}
We model a one-dimensional IAW problem via the waterbag model \cite{Berk},
i.e. assume a constant ion phase space distribution $f(u,x,t)=\frac{1}{%
2\Delta }$ between two limiting trajectories $u_{1,2}(x,t)$ and vanishing
distribution outside these trajectories \cite{Lazar142}. The distribution
remains\ constant between and outside the limiting trajectories as they are
deformed in the driven problem and, thus, the waterbag dynamics is governed
by the following dimensionless momentum and Poisson equations%
\begin{gather}
\partial _{t}u_{1}+u_{1}\partial _{x}u_{1}=-\partial _{x}\phi ,  \label{4} \\
\partial _{t}u_{2}+u_{2}\partial _{x}u_{2}=-\partial _{x}\phi ,  \label{5} \\
\partial _{x}^{2}\phi =e^{\phi +\phi _{d}}-(u_{1}-u_{2})/2\Delta .  \label{7}
\end{gather}%
Here we use time, position and ion fluid velocity $u$ normalized with
respect to the inverse ion plasma frequency $\omega _{pi}^{-1}=\left(
m_{i}/m_{e}\right) ^{1/2}\omega _{p}^{-1}$, the Debye length $\lambda
_{D}=u_{e}/\omega _{p}$, and the modified electron thermal velocity $\left(
m_{e}/m_{i}\right) ^{1/2}u_{e}$. The plasma density and the electric
potential are normalized with respect to the unperturbed plasma density and $%
k_{B}T_{e}/e$, respectively, while $2\Delta $ measures the initial velocity
width of the ion distribution. We also use the assumption of maxwellian
electrons, while the driving (ponderomotive) potential is $\phi
_{d}=\varepsilon \cos \theta _{d}$, $\varepsilon \ll 1$, and $\theta
_{d}=kx-\int \omega _{d}dt$, where the driving frequency is slowly
increasing in time, $\omega _{d}=\omega _{0}+\alpha t$. Equations (\ref{4})-(%
\ref{7}) are equivalent to a more conventional fluid system
\begin{gather}
\partial _{t}n+\partial _{x}(un)=0,  \label{1} \\
\partial _{t}u+uu_{x}=-\partial _{x}\phi -3\sigma ^{2}n\partial _{x}n,
\label{2} \\
\partial _{x}^{2}\phi =\exp (\phi +\phi _{d})-n.  \label{3}
\end{gather}%
with adiabatic ion pressure scaling $p\sim n^{3}$and $\sigma ^{2}=T_{i}/T_{e}
$ being the ratio of the ion and electron temperatures. The transition
between the two models is accomplished by setting\ $\Delta ^{2}=3\sigma ^{2}$
and relating
\begin{gather}
u=(u_{1}+u_{2})/2,  \label{8} \\
n=(u_{1}-u_{2})/2\Delta .  \label{9}
\end{gather}%
\begin{figure}[bp]
\centering \includegraphics[width=8.8cm]{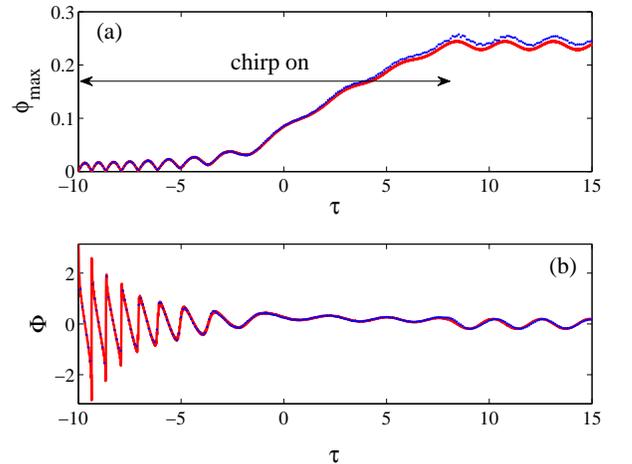}
\caption{(Color online) The maximum of the ion acoustic wave potential $%
\protect\phi _{max}$ [panel (a)] and the phase mismatch between the driven
and driving waves [panel (b)] versus slow time $\protect\tau $. The drive is
still on after $\protect\tau =8$, but its frequency chirp is off. The
waterbag model is represented by full red lines and Vlasov-Poisson
simulations are shown by dotted blue lines.}
\label{fig2}
\end{figure}
We illustrate the excitation of a chirped-driven IAW by solving Eqs. (\ref{4}%
)-(\ref{7}) numerically using a standard pseudospectral method \cite{Canuto}%
, subject to uniform initial and periodic boundary conditions $%
u_{1,2}(x,0)=\pm \Delta $, $\phi (x,0)=0$, $u_{1,2}(x+L,t)=u_{1,2}(x,t)$, $%
\phi (x+L,t)=\phi (x,t),$ $L=2\pi /k$ and parameters $\sigma =0.05$, $k=0.5$%
, $\varepsilon =0.008$, $\alpha =0.0002$. The driving frequency $\omega
_{d}=\omega _{0}+\alpha t$ is swept from below through the linear IAW wave
frequency
\begin{equation}
\omega _{0}=k\sqrt{\Delta ^{2}+\frac{1}{1+k^{2}}}  \label{10}
\end{equation}%
in the problem. Figure 1 shows the spatial distribution of the density $n$
[panel (a)] and potential $\phi $ [panel (b)] of the wave (in the wave
frame) at three \ values of slow time $\tau =\sqrt{\alpha }t$, i.e. $\tau =0$
(at the linear resonance), $\tau =4$, and $\tau =8$ by starting at $\tau =-10
$. Note that the density develops a sharply peaked spatial profile, with the
maximum ion density at $\tau =8$ reaching nearly twice the unperturbed
density ($n_{0}=1$). Additional details of the excitation process are shown
in Fig. 2, where panel (a) presents the time evolution of the maximum value
of the density wave, while panel (b) shows the evolution of the phase
mismatch \ between the driven and driving waves. The driving frequency in
Figs. 1 and 2 is chirped linearly in time until $\tau =8$ and remains
constant for $\tau >8$. One can see that the system phase locks in passing
the linear resonance and that the phase locking continues (the system is in
autoresonance), while the wave amplitude increases in average during the
chirped stage of excitation. The phase locking also persists after the chirp
is switched off and at all stages of excitation both the wave amplitude and
the phase mismatch exhibit slow oscillating modulations around the average,
indicating modulational stability in the problem. We have also compared the
results of our simulations of Eqs. (\ref{4})-(\ref{7}) with full kinetic
simulations of the associated Vlasov-Poisson system \cite{Lazar142}
\begin{figure}[bp]
\centering \includegraphics[width=8.8cm]{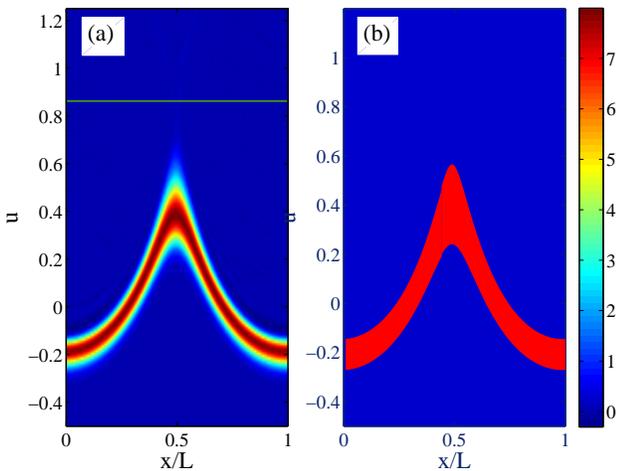}
\caption{(Color online) The ion phase space distribution at the final time, $%
\protect\tau =15$, in the example in Figs. 1 and 2; (a) The full kinetic
simulations and (b) the waterbag model. The horizontal line in panel (a)
shows the final location of the phase velocity of the driving wave.}
\label{fig3}
\end{figure}
\begin{equation}
f_{t}+uf_{x}+\phi _{x}f_{u}=0,\text{ }\phi _{xx}=\exp (\phi +\phi _{d})-\int
fdu,  \label{11}
\end{equation}%
where initially $\phi (x,0)=0$ and $f(x,u,0)=(2\pi \sigma ^{2})^{-1/2}\exp
(-u^{2}/2\sigma ^{2})$. The results of these simulations is shown by dots in
Fig. 2, showing excellent agreement with the waterbag model, while the
actual ion phase space distribution $f(x,u,\tau )$ associated with the
excited IAW in this example at $\tau =8$ is shown in Fig. 3a. For
comparison, Fig. 3b shows the corresponding waterbag distribution. The
Figure illustrates the proximity to the kinetic wave breaking limit at this $%
\tau $, as some particles in the spatial peak of the distribution have
velocities close to the phase velocity ($v_{d}=0.709$) of the driving wave.
We find numerically, that if the chirp of the driving frequency is continued
beyond $\tau =8$, the phase locking and stability of the excited wave are
destroyed. Finally, we have also tried a driving protocol (not illustrated
in the Figures), where after the autoresonant excitation stage, instead of
terminating the driving frequency chirp at $\tau =8$, we set $\varepsilon =0$%
, i.e. switched the drive off for $\tau >8$ . The result was a free, stable
large amplitude IAW with frequency close to the final frequency of the
drive. Next, we proceed to the theory of chirped-driven IAWs.

\section{Fully nonlinear theory of autoresonant ion-acoustic waves}

Our theory is based on the waterbag model (\ref{4})-(\ref{7}), where we
define potentials $\psi _{1,2}$ via $u_{1,2}=\partial _{x}\psi _{1,2}$ and
assume a weak drive:%
\begin{gather}
\partial _{tx}^{2}\psi _{1}+\partial _{x}\psi _{1}\partial _{xx}^{2}\psi
_{1}=-\partial _{x}\phi ,  \notag \\
\partial _{tx}^{2}\psi _{2}+\psi _{2x}\partial _{xx}^{2}\psi _{2}=-\partial
_{x}\phi ,  \label{system} \\
\partial _{xx}^{2}\phi =e^{\phi }(1+\phi _{d})-(\partial _{x}\psi
_{1}-\partial _{x}\psi _{2})/2\Delta .  \notag
\end{gather}%
This system can be obtained via the variation principle $\delta (\int
Ldxdt)=0$, with the three-field Lagrangian density $L=L_{0}+L_{1}$, where
\begin{eqnarray}
L_{0} &=&\frac{1}{2}(\partial _{x}\phi )^{2}+e^{\phi }-\frac{(\partial
_{x}\psi _{1}\partial _{t}\psi _{1}-\partial _{x}\psi _{2}\partial _{t}\psi
_{2})}{4\Delta }  \label{1a} \\
&&-\frac{(\partial _{x}\psi _{1})^{3}-(\partial _{x}\psi _{2})^{3}}{12\Delta
}-\frac{(\partial _{x}\psi _{1}-\partial _{x}\psi _{2})\phi }{2\Delta }
\notag
\end{eqnarray}%
and $L_{1}=e^{\phi }\phi _{d}$. We seek solutions of form $\phi =\phi
(\theta )$, $\psi _{1,2}=\beta _{1,2}x+V_{1,2}(\theta )$, where $\phi
(\theta )$ and $V_{1,2}(\theta )$ are $2\pi $-periodic in\ fast phase $%
\theta =kx-\int \Omega dt$, while $\beta _{1,2}$ are constant and $\Omega $
is slow frequency reflecting the slow frequency chirp of the drive. We also
assume that $\phi (\theta )\,\ $and $\partial _{x}V_{1,2}(\theta )$ have
zero $\theta $-averages, yielding $\beta _{1,2}=\pm \Delta $. Next, we
replace $\partial _{x}\psi _{1,2}=\beta _{1,2}+\partial _{x}V_{1,2}$ and $%
\partial _{t}\psi _{1,2}=-v_{p}\partial _{x}V_{1,2}$ in the Lagrangian
density ($v_{p}=\Omega /k$ being the wave phase velocity) and drop the terms
which do not include field variables. This yields the unperturbed Lagrangian
density

\begin{eqnarray}
L_{0} &=&\frac{1}{2}(\partial _{x}\phi )^{2}-\phi +e^{\phi }-\frac{\left(
\Delta -v_{p}\right) }{4}\left[ \frac{(\partial _{x}V_{1})^{2}}{\Delta }%
+\partial _{x}V_{1}\right]   \notag \\
&&-\frac{\left( \Delta +v_{p}\right) }{4}\left[ \frac{(\partial
_{x}V_{2})^{2}}{\Delta }+\partial _{x}V_{2}\right]   \label{4a} \\
&&-\frac{(\partial _{x}V_{1})^{3}-(\partial _{x}V_{2})^{3}}{12\Delta }-\frac{%
(\partial _{x}V_{1}-\partial _{x}V_{2})\phi }{2\Delta }.  \notag
\end{eqnarray}%
For fixed $v_{p}$, the problem described by $L_{0}$ is integrable. Indeed,
we have two constant canonical momenta $p_{1,2}=\partial L_{0}/\partial
(\partial _{x}V_{1,2})$%
\begin{eqnarray}
p_{1} &=&-\frac{\phi }{2\Delta }-\frac{\left( \Delta -v_{p}\right) }{4}%
\left( \frac{2\partial _{x}V_{1}}{\Delta }+1\right) -\frac{(\partial
_{x}V_{1})^{2}}{4\Delta },  \label{5a} \\
p_{2} &=&\frac{\phi }{2\Delta }-\frac{\left( \Delta +v_{p}\right) }{4}\left(
\frac{2\partial _{x}V_{2}}{\Delta }-1\right) +\frac{(\partial _{x}V_{2})^{2}%
}{4\Delta },  \label{6a}
\end{eqnarray}%
which can be used to express%
\begin{eqnarray}
\partial _{x}V_{1} &=&v_{p}-\Delta -s_{1},  \label{7a} \\
\partial _{x}V_{2} &=&v_{p}+\Delta -s_{2},  \label{8a}
\end{eqnarray}%
where%
\begin{equation}
s_{1,2}=\sqrt{2(B_{1,2}-\phi )}.  \label{9a}
\end{equation}%
In the definition of $s_{1,2}$, we use new conserved parameters, $B_{1,2}$,
instead of $p_{1,2}$:%
\begin{eqnarray}
B_{1} &=&\frac{v_{p}}{2}(v_{p}-\Delta )-2p_{1}\Delta ,  \label{10a} \\
B_{2} &=&\frac{v_{p}}{2}(v_{p}+\Delta )+2p_{2}\Delta .  \label{11a}
\end{eqnarray}%
Note that initially ($\phi =\partial _{x}V_{1}=\partial _{x}V_{2}=0$), $%
p_{1}=(v_{p}-\Delta )/4$, $p_{2}=(v_{p}+\Delta )/4$, $B_{1}=(v_{p}-\Delta
)^{2}/2$, and $B_{2}=(v_{p}+\Delta )^{2}/2$. The choice of the signs at $%
s_{1,2}$ in (\ref{7a}) and (\ref{8a}) is such that $\partial _{x}V_{1,2}=0$
initially. In addition to $B_{1,2}$, the energy function
\begin{equation}
A^{\prime }=(\partial _{x}\phi )^{2}+p_{1}\partial _{x}V_{1}+p_{2}\partial
_{x}V_{2}-L_{0}  \label{12}
\end{equation}%
is also conserved in the fixed $v_{p}$ case. Then, by using (\ref{10a}), and
(\ref{11a}) in (\ref{12}), we get the usual energy conservation-type
equation
\begin{equation}
\frac{1}{2}\phi _{x}^{2}+U_{eff}=A  \label{14}
\end{equation}%
where the effective potential is%
\begin{eqnarray}
U_{eff} &=&-\frac{1}{2}v_{p}^{2}-\frac{B_{1}}{2\Delta }(v_{p}-\Delta )+\frac{%
B_{2}}{2\Delta }(v_{p}+\Delta )+\frac{s_{1}^{3}-s_{2}^{3}}{6\Delta }  \notag
\\
&&-e^{\phi }+1-\frac{\Delta ^{2}}{6}  \label{15}
\end{eqnarray}%
and $A=A^{\prime }+1-\Delta ^{2}/6$. We have added $1-\Delta ^{2}/6$ to $%
A^{\prime }$ to make the effective potential zero for the initially
unperturbed ($\phi =0$) plasma. Indeed, initially, $p_{1}=(v_{p}-\Delta )/4,$
$p_{2}=(v_{p}+\Delta )/4$ and $B_{1}=(v_{p}-\Delta )^{2}/2$, $%
B_{2}=(v_{p}+\Delta )^{2}/2$ and by expanding $U_{eff}$ in $\phi $ to second
order we get%
\begin{equation}
U_{eff}=\frac{1}{2}\left( 1-\frac{1}{v_{p}^{2}-\Delta ^{2}}\right) \phi
^{2}+O(\phi ^{3}).  \label{SmallPhi}
\end{equation}%
Then the spatial frequency (i.e., $k$) of small oscillations of $\phi $ is
\begin{equation}
k=\sqrt{\frac{1}{v_{p}^{2}-\Delta ^{2}}-1}  \label{k}
\end{equation}%
in agreement with the linear dispersion relation (\ref{10}). Next, following
Whitham's procedure \cite{Williams}, we average (\ref{12}) over $\theta $ to
obtain the averaged Lagrangian density $\Lambda _{0}=\left\langle
L_{0}\right\rangle _{\theta }$:%
\begin{gather}
\Lambda _{0}(A,B_{1},B_{2};v_{p})-1+\Delta ^{2}/6=\left\langle (\partial
_{x}\phi )^{2}\right\rangle -A  \label{16} \\
=kI(A,B_{1},B_{2};v_{p})-A,  \notag
\end{gather}%
where
\begin{equation}
I=\frac{1}{2\pi k}\int_{0}^{2\pi }\phi _{x}^{2}d\theta =\frac{1}{2\pi }\oint
[2(A-U_{eff})]^{1/2}d\phi .  \label{17}
\end{equation}%
is the usual action integral. The perturbed part of the averaged Lagrangian
density is%
\begin{equation}
\Lambda _{1}=\left\langle e^{\phi }\phi _{d}\right\rangle _{\theta }=\frac{%
\varepsilon }{2}a_{1}(I)\cos \Phi ,  \label{18}
\end{equation}%
where we have expanded $e^{\phi }=\sum a_{n}(I)\cos (n\theta )$, neglected
all but fundamental harmonic in this expansion (this is the isolated
resonance approximation \cite{Chirikov}), wrote $\theta _{d}=\theta -\Phi $,
and assumed slow phase mismatch $\Phi $ in the problem. Then, the full
averaged Lagrangian density is%
\begin{equation}
\Lambda =kI(A,B_{1},B_{2};v_{p})-A+\frac{\varepsilon }{2}a_{1}(I)\cos \Phi .
\label{19}
\end{equation}%
This Lagrangian density can be used by taking variations with respect $%
A,B_{1},B_{2}$, and $\theta $ to yield%
\begin{gather}
k\partial _{A}I-1+\frac{\varepsilon }{2}\partial _{A}a_{1}\cos \Phi =0,
\label{20} \\
k\partial _{B_{1}}I+\frac{\varepsilon }{2}\partial _{B_{1}}a_{1}\cos \Phi =0,
\label{21} \\
k\partial _{B_{2}}I+\frac{\varepsilon }{2}\partial _{B_{2}}a_{1}\cos \Phi =0,
\label{22}
\end{gather}%
and%
\begin{equation}
\frac{d}{dt}\left( \partial _{v_{p}}I\right) =\frac{\varepsilon }{2}%
a_{1}(I)\sin \Phi .  \label{23}
\end{equation}%
With the addition of
\begin{equation}
\frac{d\Phi }{dt}=\Omega -\omega _{d}(t),  \label{24}
\end{equation}%
we now have a complete system of slow equations for $A,B_{1},B_{2},\Omega ,$
and $\Phi $. The phase-locked quasi-equilibrium $\Phi \approx 0$ in this
system is obtained via solving a simpler set of three algebraic equations
for $A,B_{1},B_{2}$
\begin{gather}
k\partial _{A}I-1\approx 0,  \label{25} \\
\partial _{B_{1}}I\approx 0,  \label{26} \\
\partial _{B_{2}}I\approx 0,  \label{27}
\end{gather}%
where $\Omega =\omega _{d}(t).$

To check our theory in a simplified problem, take the limit $\Delta
\rightarrow 0$ and consider the undriven case, $v_{p}$ being the phase
velocity of the undriven wave. Furthermore, assume $B_{1}=(v_{p}-\Delta
)^{2}/2$ and $B_{2}=(v_{p}+\Delta )^{2}/2$, as for the linear equilibrium.
In this case, $(B_{1}+B_{2})/2\rightarrow v_{p}^{2}/2$, $%
v_{p}(B_{1}-B_{2})/2\Delta \rightarrow -v_{p}^{2}$, and%
\begin{equation}
(s_{1}^{3}-s_{2}^{3})/6\Delta \rightarrow \frac{3s^{2}}{6}\frac{s_{1}-s_{2}}{%
B_{1}-B_{2}}\frac{B_{1}-B_{2}}{\Delta }\rightarrow -\frac{v_{p}s^{2}}{2}%
\frac{\partial s}{\partial B}=-v_{p}s,  \label{30}
\end{equation}%
where $s=\sqrt{v_{p}^{2}-2\phi }$. Then Eq. (\ref{15}) yields the well known
Sagdeev potential for the IAWs \cite{FFChen}:
\begin{equation}
U_{eff}\rightarrow v_{p}^{2}-v_{p}\sqrt{v_{p}^{2}-2\phi }-e^{\phi }+1.
\label{31}
\end{equation}%
\begin{figure}[bp]
\centering \includegraphics[width=8.8cm]{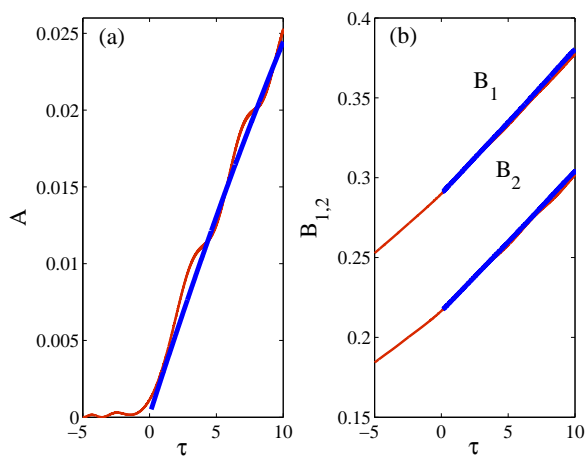}
\caption{(Color online) An example of the time evolution of the quasi-energy
$A$ and slow parameters $B_{1,2}$ from the averaged Lagrangian theory (thick
blue lines) and from the full fluid simulations (thinner red lines).}
\label{fig4}
\end{figure}
\begin{figure}[tp]
\centering \includegraphics[width=8.8cm]{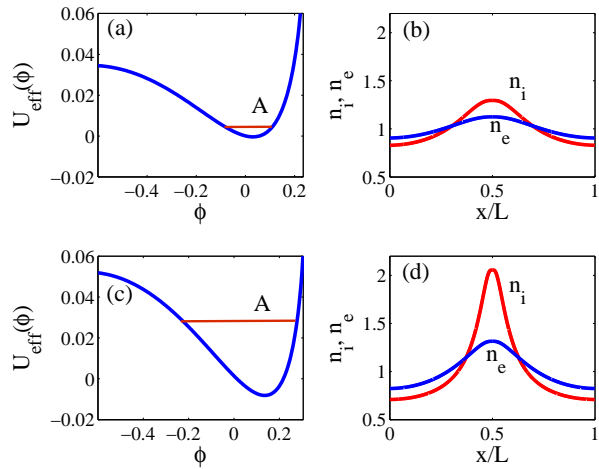}
\caption{(Color online) The effective potential $U_{eff}(\protect\phi )$ and
the spatial distribution of the densities of plasma species $n_{i,e}$ from
the averaged Lagrangian theory at two times $\protect\tau =2$ (panels (a)
and (b)) and $\protect\tau =10$ (panels (c) and (d)). The parameters in
these examples correspond to those of Fig. 4 and the horizontal red lines in
panels (a) and (c) show the value of the quasi-energy $A$ at the
corresponding times.}
\label{fig5}
\end{figure}
In this case, $I=I(A,v_{p})$ and Eq. (\ref{25}) yields the nonlinear
dispersion relation $v_{p}=v_{p}(A)$ of the wave. In contrast to this well
known problem, in the chirped autoresonant case, we cannot assume that $%
B_{1,2}$ and $A$ remain constant during the evolution. In the quasi-static
approximation, these slow variables are described by Eqs. (\ref{25})-(\ref%
{27}) for a given dependence of the driving frequency $\omega _{d}\ $on
time. We present an example of such calculations in the case of $\sigma =0.03
$ and driving parameters $k=1$, $\varepsilon =0.008$, and $\alpha =0.0001$.
Figure 4 shows (thick lines) the evolution of the quasi-energy $A$ and slow
parameters $B_{1,2}$ obtained by solving the algebraic system (\ref{25})-(%
\ref{27}) numerically. In the same Figure, the numerical results from the
full fluid system (\ref{4})-(\ref{7}) are shown by thin lines. One observes
a good agreement beyond the linear resonance, the discrepancy due to the
neglect of $\varepsilon $ in (\ref{25})-(\ref{27}). One can also see
oscillating modulations of $A~\ $and $B_{1,2}$ in fluid simulations around
the quasi-equilibrium. As mentioned earlier, these modulations are
characteristic of many autoresonant problems and reflect modulational
stability of the autoresonant evolution. Their detailed analysis requires
numerical solution of Eqs. (\ref{20})-(\ref{24}), which is beyond the scope
of this paper. Additional results from the variational theory in our example
are presented in Fig. 5, which shows the quasi-potential $U_{eff\text{ }}$
(Fig. 5a and 5c) and the ion and electron densities $n$ and $\exp (\phi )$
versus $x/L$ (Figs. 5b and 5d) for two values of the driving phase velocity $%
v_{p}=0.75$ and $0.82$. The corresponding quasi-energies $A$ for these
values of $v_{p}$ are represented by horizontal red lines in the Figure. The
development of a sharply peaked density profile in the Figure is associated
with the approach of the quasi-energy to the value at which parameter $s_{1}=%
\sqrt{2(B_{1}-\phi )}$ vanishes. As mentioned earlier, the kinetic effects
limit this development. This completes the discussion of our averaged
variational approach to autoresonant IAWs and we proceed to conclusions.

\section{Conclusions}

We have studied excitation and control of large amplitude IAWs by a chirped
frequency driving wave. The process involved passage through linear
resonance in the problem and transition to autoresonant stage of excitation,
where the driven IAW self-adjusted its parameters (both it's amplitude and
frequency increased) to stay in a continuous resonance with the drive. The
method allowed reaching extreme excitation amplitudes as the ion density
developed a sharply peaked spatial profile with the maximum exceeding the
unperturbed ion density significantly. At later stages of excitation, when
the local maximum of the ion fluid velocity approached the phase velocity of
the driving wave, the autoresonant process discontinued due to the kinetic
wave breaking. These predictions were confirmed in numerical simulations
using the waterbag model [see Eqs. (\ref{4})-(\ref{7})] and compared with
fully kinetic Vlasov-Poisson simulations [Eqs. (\ref{11})]. We have also
developed the adiabatic theory for studying the formation of autoresonant
IAWs. The theory used Whitham's averaged Lagrangian approach applied to the
waterbag model. It allowed interpretation of the driven IAWs as a dynamical
problem of an oscillation of a quasi-particle in a slowly evolving effective
potential (the generalization of the Sagdeev potential). The evolution of
the energy of the quasi-particle and other slow parameters of this dynamics
can be found by solving adiabatic Eqs. (\ref{20})-(\ref{24}). We have
applied this theory in studying the quasi-static evolution of the IAWs in
the problem [this case reduces to solving algebraic Eqs. (\ref{25})-(\ref{27}%
)] and illustrated a good agreement with simulations. The averaged
Lagrangian approach is suitable for studying the stability of the extreme
IAW excitations as seen in simulations, which seems to be an important goal
for the future. We have identified the kinetic wave breaking process as
responsible for terminating the autoresonant excitation of IAWs, and
limiting the amplitude of the excited IAW. In seeking even larger amplitude
excitations one must avoid this kinetic wave breaking by decreasing the ion temperature. Alternatively, this goal can be
reached in higher $Z$ (ion charge) plasmas, where the linear ion acoustic
frequency increases by factor of $Z^{1/2}$, distancing the driving frequency
necessary for autoresonant excitation from the ion velocity distribution.
Investigation of these effects, as well as studying details of the kinetic
wave breaking process in application to autoresonant IAWs also comprise
important goals for future research.

\begin{acknowledgements}
This work was supported by the Israel Science Foundation
Grant No. 30/14
\end{acknowledgements}

\end{document}